\documentclass[aps,prl,twocolumn,superscriptaddress,showpacs,showkeys]{revtex4}
\usepackage{epsfig}
\usepackage{graphics,amsmath,amssymb}
\usepackage[english]{babel}

\begin{document}
\title{Finite size  scaling theory for percolation phase transition}
\author{Yong Zhu}
\author{Xiaosong Chen}
\affiliation{State Key Laboratory of Theoretical Physics, Institute of Theoretical Physics, Chinese Academy of Sciences, P.O. Box 2735, Beijing 100190, China}

\date{\today}

\begin{abstract}
The finite-size scaling theory for continuous phase transition plays an important role in determining critical point and critical exponents from the size-dependent behaviors of quantities in the thermodynamic limit. For percolation phase transition, the finite-size scaling form for the reduced size of largest cluster has been extended to cluster ranked $R$. However, this is invalid for explosive percolation as our results show. Besides, the behaviors of largest increase of largest cluster induced by adding single link or node have also been used to investigate the critical properties of percolation and several new exponents $\beta_1$, $\beta_2$, $1/\nu_1$ and $1/\nu_2$ are defined while their relation with $\beta/\nu$ and $1/\nu$ is unknown. Through the analysis of asymptotic properties of size jump behaviors, we obtain correct critical exponents and develop a new approach to finite size scaling theory where sizes of ranked clusters are averaged at same distances from the sample-dependent pseudo-critical point in each realization rather than averaging at same value of control parameter.
\end{abstract}

\maketitle

{\it Introduction}---
Percolation, as one of the fundamental models in statistical and condensed matter physics, was widely believed to
be a typical continuous phase transition for different space dimensions and various networks architectures\cite{Stauffer94,Dorogovtsev08}. Therefore, the so-called explosive percolation (EP) which was proposed and reported to be discontinuous by Achlioptas {\it et al}.\cite{Achlioptas} has attracted much attention and been extensively studied ever since\cite{Bastas14}. For EP under product rule (PR)\cite{Achlioptas}, instead of adding bonds one by one randomly in the original Erd\"os-R\'enyi (ER) model\cite{ER}, at each step two randomly-picked unoccupied bonds are considered and only the one with smaller product of the sizes of the two clusters to be connected will be added while the other one is discarded. This slight modification of growth procedure leads to so significant change of the properties of percolation as to arouse much popular interest. After all, EP was proved both analytically\cite{Riordan11,Costa10,Costa14} and numerically\cite{Lee11,Grassberger11,Liu11,Fan12,Bastas11,Zhu14,Tian12} to be indeed continuous phase transition with critical exponents different from ordinary percolation.

In the study of phase transition and critical phenomena, Monte Carlo simulation along with finite size scaling (FSS) theory has always been one of the most important tools since most models could not be solved analytically. %As we know, phase transition is defined in the thermodynamic limit while
FSS theory describes a build-up of the bulk properties when a small system is increased in size. For continuous percolation phase transition, the order parameter which is usually defined as $s_1(r,N)$ the reduced size of the largest cluster follows a finite size scaling form\cite{privman1,privman2}
\begin{equation}
s_1(r,N)=N^{-\beta/\nu}\tilde{s}_1(tN^{1/\nu}). \label{eq1}
\end{equation}
where the controlling parameter $r$ denotes the number of added edges divided by the system size $N$ and $t=(r-r_c)/r_c$ is the reduced deviation from the critical point $r_c$. The critical exponent $\nu$ characterizes the divergence of correlation length $\xi= \xi_0 |t|^{-\nu}$. The finite size scaling behaviour is valid in the asymptotic critical region with $N\gg 1$ and $|t|\ll 1$. Obviously, $\beta/\nu$ can be determined from the size-dependent behaviors of $s_1$ right at the critical point $r_c$.

In addition, the distribution $P(s,L)$ of order parameter at critical point $r_c$ satisfies a analogous finite size scaling hypothesis\cite{Binder81,Bruce92}
\begin{equation}
P_{r=r_c}(s,N)=N^{\beta/\nu}\tilde{P}(sN^{\beta/\nu}). \label{distribution_1}
\end{equation}
Despite the nature of continuous phase transition, EP shows unusual finite size behaviors. Rather than single humped as in ordinary percolation, $P_{r=r_c}(s,N)$ is double humped in various EP models\cite{Grassberger11,Tian12}. Although the distance between the two peaks decreases with system size $N$ in power-law, data collapse can be roughly achieved for each peak separately with different exponents $\beta/\nu$. Double-humped distribution of order parameter is also found at the pseudo-critical point $r_c(N)$ where the average cluster size reaches its peak\cite{Rad10}.

{\it Models and Method}---
We carry out extensive Monte Carlo (MC) simulation of ER model~\cite{ER}, PR model~\cite{Achlioptas}, CDGM model~\cite{Costa10} and track the cluster information with the effective algorithm of Newmann and Ziff~\cite{Newmann00,Newmann01}. In each realization of networks, we start with $N$ isolated nodes and then edges are added according to corresponding rules.

The critical point of ER model is known as $r_c=0.5$ and critical exponents $\beta/\nu=1/\nu=1/3$. PR model has been extensively studied with finite size scaling theory in Ref.~\cite{Fan12} and it is reported that
\begin{equation}
r_c=0.88845(5),1/\nu=0.5(1),\beta/\nu=0.04(1). \label{eq9}
\end{equation}
For the CDGM model, according to the detailed analysis by Costa et al.~\cite{Costa10,Costa14}, we have
\begin{equation}
r_c=0.9232075...,1/\nu=0.818(1),\beta=0.0555(1). \label{eq10}
\end{equation}
Then we can get $\beta/\nu=0.0454(2)$.

As has been demonstrated~\cite{Jan98,Jan99,Macleod98}, the finite size scaling form of $s_1(r,N)$ also stands for the sizes of second, third...larest cluster. To be explicit, for the cluster ranked $R$, we have
\begin{equation}
s_R(r,N)=N^{-\beta/\nu}\tilde{s}_R(tN^{1/\nu}). \label{eq4}
\end{equation}
Therefore, all the sizes of ranked clusters scale as $s_R\propto N^{-\beta/\nu}$ at the critical point $r_c$. Besides, the intersection of curves $s_2/s_1$ with different system sizes $N$ can be used to estimate critical point~\cite{Liu11,Fan12,Zhu14,Silva00,Silva02,Alb05} since it is a $N$-independent constant at critical point.

\begin{figure}
\resizebox{0.45\textwidth}{!}{\includegraphics{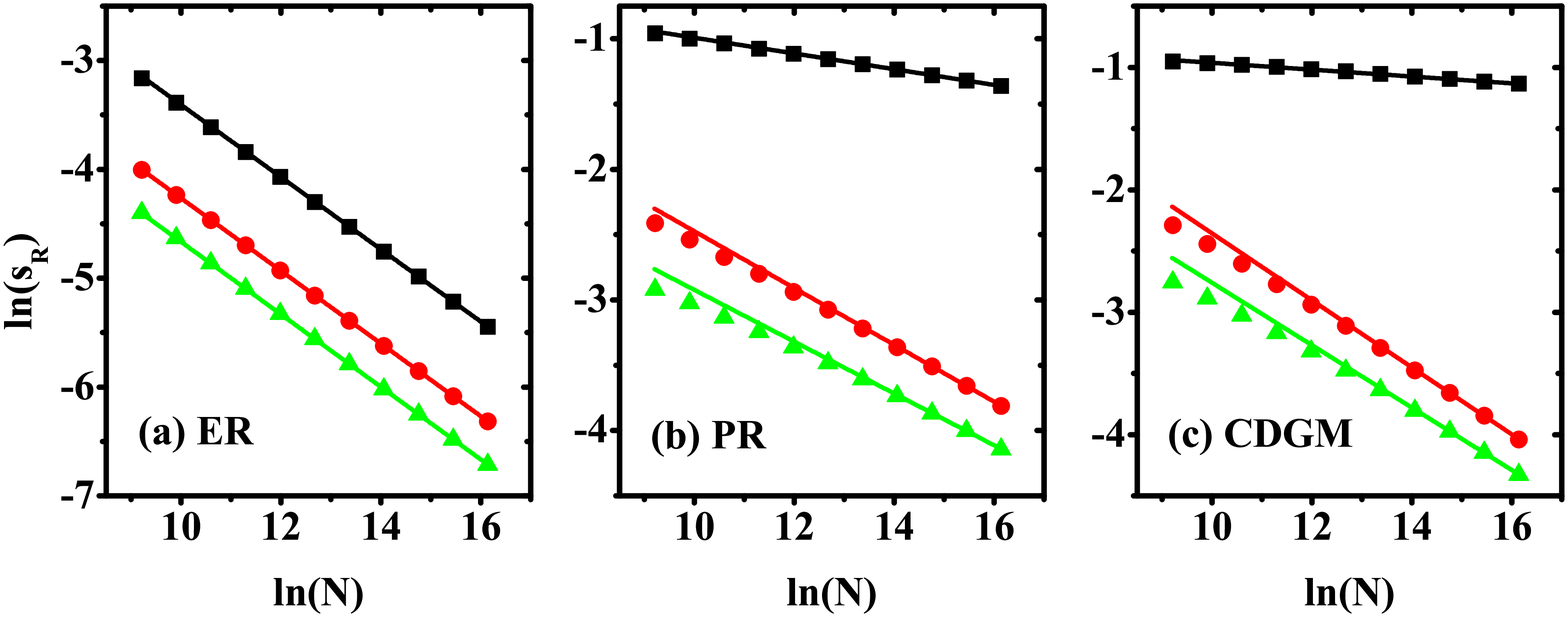}}
\caption{Plots of $\ln{s_R}$ versus $\ln{N}$ with $R=1$ ($\square$, black), 2 ($\circ$, red), 3 ($\triangle$, green) at critical point. The solid lines represent the linear fitting of data with three largest system sizes. (A)ER model at $r_c=0.5$. (B)PR model at $r_c=0.888444$. (C)CDGM model at $r_c=0.9232075$. Critical exponent ratios $\beta/\nu$ estimated from $\ln{s_1}$, $\ln{s_2}$ and $\ln{s_3}$ are summarized in Table~\ref{table1}.}
\label{fig:1}       % Give a unique label
\end{figure}

In Fig.\ref{fig:1}(a), $\ln{s_R}$ with respect to $\ln{N}$ at the critical point for the ER model is shown. We see excellent and parallel straight lines and the critical exponent ratios $\beta/\nu$ estimated from $\ln{s_1}$, $\ln{s_2}$ and $\ln{s_3}$ are respectively $0.332(2)$, $0.334(1)$ and $0.333(1)$ which are equal to each other within error bar and agree quite well with the analytic value mentioned above.

However, it's quite different for explosive percolation transitions as shown in Fig.\ref{fig:1}(b) and (c) for PR and CDGM model respectively. It's clear that $s_2$ and $s_3$ decrease much faster than $s_1$ with the increase of $N$ and obvious finite size effects appear. Through the linear fitting of data with system size $N/10^4=256, 512, 1024$, the critical exponent ratios $\beta/\nu$ for the three largest clusters are estimated as $0.060(2)$, $0.22(2)$ and $0.20(2)$ for PR model and $0.0266(1)$, $0.275(5)$ and $0.259(6)$ for CDGM model respectively. As a consequence, $s_2/s_1$ at $r_c$ turns to be $N$-dependent in EP. Furthermore, for CDGM model, none of the obtained $\beta/\nu$ agrees with previous analytical result $\beta/\nu=0.0454(2)$.

Apart from the behaviors of the average value of ranked clusters, percolation phase transitions can also be characterized by the size jump behaviors of order parameter in each realization of networks\cite{Lee11,Nagler11,Qian12,Fan14a,Fan14b,Zhu14}. Here, we define sample-dependent pseudo-critical point as the $r_c^{i}$ where order parameter exhibits a sudden biggest jump in $i$-th realization and the corresponding jump gap is denoted as $\Delta^{(i)}$. For a network of each size $N$, $M=1024,000$ realizations of network are made. From the results of all simulations, we can calculate the averages
\begin{eqnarray}
\bar{\Delta}(N)&= \frac{1}{M}\sum_{i=1}^{M} \Delta^{(i)},\\
\bar{r}_c (N) &= \frac{1}{M}\sum_{i=1}^{M} r_c^{(i)}.
\end{eqnarray}
and root mean squares of fluctuations $\delta r_c = r_c - \bar{r}_c (N)$ and $\delta \Delta = \Delta - \bar{\Delta}(N)$
\begin{eqnarray}
\chi_r &\equiv \sqrt{<(\delta r_c)^2>}, \\
\chi_\Delta &\equiv \sqrt{<(\delta \Delta)^2>}.
\end{eqnarray}

The following finite size scaling hypotheses are made and confirmed by simulation data for continuous percolation~\cite{Zhu14,Fan14a,Fan14b}
\begin{align}
\bar{r}_c (N)&= r_c (\infty)+a_r N^{-1/\nu_1}, \label{eq_r_c} \\
\bar{\Delta} (N)&= a_\Delta N^{-\beta_1},  \label{eq_delta}\\
\chi_r &= b_r N^{-1/\nu_2},   \label{eq_chi_r}\\
\chi_\Delta &= b_{\Delta} N^{-\beta_2}. \label{eq_chi_delta}
\end{align}

We anticipate that not only the largest but also the second, third... largest jump of order parameter $s_1$ and the corresponding sample-dependent pseudo-critical points are related to percolation phase transition and satisfy the finite size scaling hypotheses above. Denote $\bar{r}_c^k(N)$ as the average sample-dependent pseudo-critical point where the $k$th largest jump of $s_1$ occurs, then it's easy to get
\begin{equation}
\bar{r}_c^{k_1}(N)-\bar{r}_c^{k_2}(N)= c N^{-1/\nu_1}. \label{eq_dr_c}
\end{equation}
It provides a more reliable way to estimated $1/\nu_1$ because $1/\nu_1$ is quite sensitive to the value of $r_c (\infty)$ when fitting with Eq.\eqref{eq_r_c}.
%Furthermore, critical exponents $1/\nu_1$ and $1/\nu_2$ equal to $1/\nu$ characterizing the divergence of correlation length while $\beta_1$ and $\beta_2$ equal to $\beta/\nu$ characterizing the convergence of order parameter. If $\bar{\Delta} (N)$ converges to a nonzero constant as $N \to \infty$, the jump of order parameter will not disappear and the percolation phase transition turns out to be discontinuous.

\begin{figure}
\resizebox{0.5\textwidth}{!}{\includegraphics{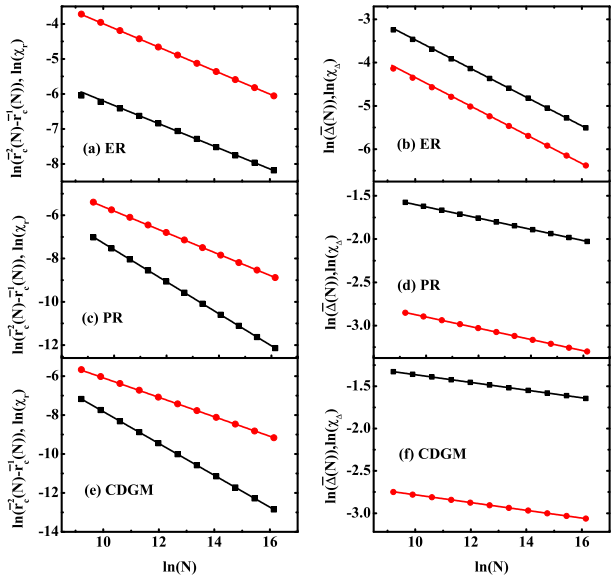}}
\caption{Plots of $\ln(\bar{r}_c^2(N)-\bar{r}_c^1(N))$, $\ln(\chi_r)$, $\ln(\bar{\Delta}(N))$ and $\ln(\chi_\Delta)$ versus $\ln N$ for ER, PR and CDGM model. For the sake of comparison, data of average value ($\square$, black) and the root mean squares of its fluctuation ($\circ$, red) are shown together in each panel. Critical exponent ratios $\beta_1$, $\beta_2$, $1/\nu_1$ and $1/\nu_2$ are summarized in Table~\ref{table1}.}
\label{fig:2}       % Give a unique label
\end{figure}

{\it Results of MC simulation}---
The asymptotic properties of the above four quantities associated with the size jump behaviors of order parameter are shown in Fig.\ref{fig:2}. For the sake of comparison, data of $\ln(\bar{r}_c^2(N)-\bar{r}_c^1(N))$ and $\ln(\chi_r)$ are plotted together while data of $\ln(\bar{\Delta}(N))$ and $\ln(\chi_\Delta)$ are plotted together.

In Fig.\ref{fig:2}(a) and (b), we can see clearly two parallel lines in each panel for ER model. From the slope of fitting lines, we get $1/\nu_1=0.333(4)$ and $1/\nu_2=0.335(2)$ which are in consistent with $1/\nu=1/3$, along with $\beta_1=0.331(3)$ and $\beta_2=0.331(6)$ which are in consistent with $\beta/\nu=1/3$.

For PR model, fitting lines for $\ln(\bar{r}_c^2(N)-\bar{r}_c^1(N))$ and $\ln(\chi_r)$ in Fig.\ref{fig:2}(c) are apparently not parallel, indicating critical exponent $1/\nu_1$ is different from $1/\nu_2$. From their slopes, we get $1/\nu_1=0.740(1)$ and $1/\nu_2=0.501(3)$. The asymptotic behaviors of $\bar{\Delta}(N)$ and $\chi_\Delta$ are shown in Fig.\ref{fig:2}(d). The exponents $\beta_1=0.0650(2)$ and $\beta_2=0.0650(1)$ are obtained.

As shown in Fig.\ref{fig:2}(e) and (f), the situation for CDGM model is similar to PR model. Critical exponents are estimated as $1/\nu_1=0.818(1)$, $1/\nu_2=0.500(4)$, $\beta_1=0.0455(1)$ and $\beta_2=0.0458(3)$.  In comparison with the results of Costa et al.~\cite{Costa10,Costa14}, $1/\nu_1=0.818(1)$ is consistent with $1/\nu=0.818$ while $\beta_1$ and $\beta_2$ are in accordance with $\beta/\nu=0.0454(2)$. So far, from the jump behaviors of largest cluster, the inherent critical exponents $1/\nu$ and $\beta/\nu$ for CDGM model are obtained. Likewise, we infer those for PR model are $0.740$ and $0.065$. All the critical exponents obtained in this section are summarized in Table~\ref{table1}. It's worth mentioning that, with standard FSS theory, $1/\nu$ for PR model is estimated as 0.5\cite{Fan12} which is equal to $1/\nu_2$. It's not a coincidence since same phenomenon can be observed for CDGM model.

\begin{table}[!h]
\caption{Critical exponent $\beta_1$, $\beta_2$, $1/\nu_1$ and $1/\nu_2$ obtained from the size jump behaviors of largest cluster. }
\begin{tabular}{llllll}
\hline \hline
 Model& $\beta_1$ & $\beta_2$ & $1/\nu_1$ & $1/\nu_2$  \\ \hline
ER & 0.331(3) & 0.331(6) & 0.333(4) & 0.335(2) \\
PR & 0.0650(2) & 0.0650(1) & 0.740(1) &0.501(3) \\
CDGM & 0.0455(1) & 0.0458(3) & 0.818(1) & 0.500(4) \\ \hline \hline
\end{tabular}
\label{table1}
\end{table}

{\it Modified Finite Size Scaling Theory}---
In standard finite size scaling theory for percolation, an quantity $X$ is averaged over the different samples at the same value of control parameter or at the same shift from the critical point of infinite systems. For instance, the reduced sizes of ranked clusters are computed as
\begin{equation}
s(r,N)=[s^{(i)}(r,N)]=[s^{(i)}(\delta,N)]. \label{eq_s(dr,N)}
\end{equation}
where $\delta=r-r_c$ is the deviation from critical point $r_c(\infty)$, the superscript $i$ stands for the $i$th sample or realization and $[\cdots]$ stands for ensemble average.

In the field of disorder systems~\cite{Stinchcombe,Shalaev,Selke} where the disorders are usually induced by site dilution or bond-randomness, there is an additional source of fluctuation from the variation in the transition temperature itself besides thermal fluctuations. So, different measured values are obtained in every sample with different configuration of the quenched disorder. Therefore, instead of the conventional finite size scaling, a sample dependent scaled variable is proposed to be $\dot{t}^{(i)}=(T-T_c^{(i)}(N))/T_c$ where $T_c^{i}(N)$ is a pseudo-critical temperature of sample $i$. In terms of $\dot{t}^{(i)}$, a quantity $X$ after thermal averaging is expected to show a sample dependent finite size scaling form~\cite{Wiseman95,Pazmandi,Wiseman98a,Wiseman98b,Bernarder00}
\begin{equation}
X^{(i)}(T,N)=N^{\rho}Q(\dot{t}^{(i)}N^{1/\nu}).
\end{equation}
which is equivalent to the common one in Ref.\cite{Barber}. As mentioned in Ref.\cite{Wiseman98b}, $X^{(i)}$ with the same $\dot{t}^{(i)}$ and $N$ from different samples are still slightly different. Thus, whether it's necessary to do the sample averaging leads to the problem of self-averaging. If the ratio of variance $V_X=[(X^{(i)}-\bar{X}^{(i)})^2]$ and the square of shift $\delta{X}=\bar{X}^{(i)}-X(\infty)$ converges to 0 as $N \to \infty$, indicating that the distribution of $X^{(i)}$ tends to be a $\delta$-function right at $X(\infty)$, then $X$ possesses the property of self-averaging. In this case, the measurement of $X^{(i)}$ in one large sample $i$ will provide a good estimate of the sample average.

However, percolation is a typical non-self-averaging model since neither $r_c^{i}(N)$ nor $\delta_c^{i}(N)$ is self-averaging in any model we studied here according to the scaling of $\bar{r}_c(N)$, $\bar{\Delta}(N)$, $\chi_r$ and $\chi_\Delta$. So we define the reduced sizes of ranked clusters in percolation process as
\begin{equation}
C_R(\dot{\delta},N)=[C_R^{(i)}(\dot{\delta},N)]. \label{eq_C_R}
\end{equation}
where $\dot{\delta}=r-r_c^{(i)}$ is the deviation from pseudo-critical point $r_c^{(i)}$ in each realization and averaging over different realizations are made at the same $\dot{\delta}$. Then we anticipate a finite-size scaling form of $C_R$ as
\begin{equation}
C_R(\dot{\delta},N)=N^{-\beta/\nu}\tilde{C}_R(\dot{t}N^{1/\nu}). \label{eq_mffs}
\end{equation}
where $\dot{t}=\dot{\delta}/r_c(\infty)$ is the reduced deviation from sample-dependent pseudo-critical point. The critical exponent $\beta/\nu$ is inferred to be $\beta_1$ and $1/\nu$ inferred to be $1/\nu_1$ obtained from the size jump behaviors of largest cluster as summarized in Table~\ref{table1}.

\begin{figure}
\resizebox{0.5\textwidth}{!}{\includegraphics{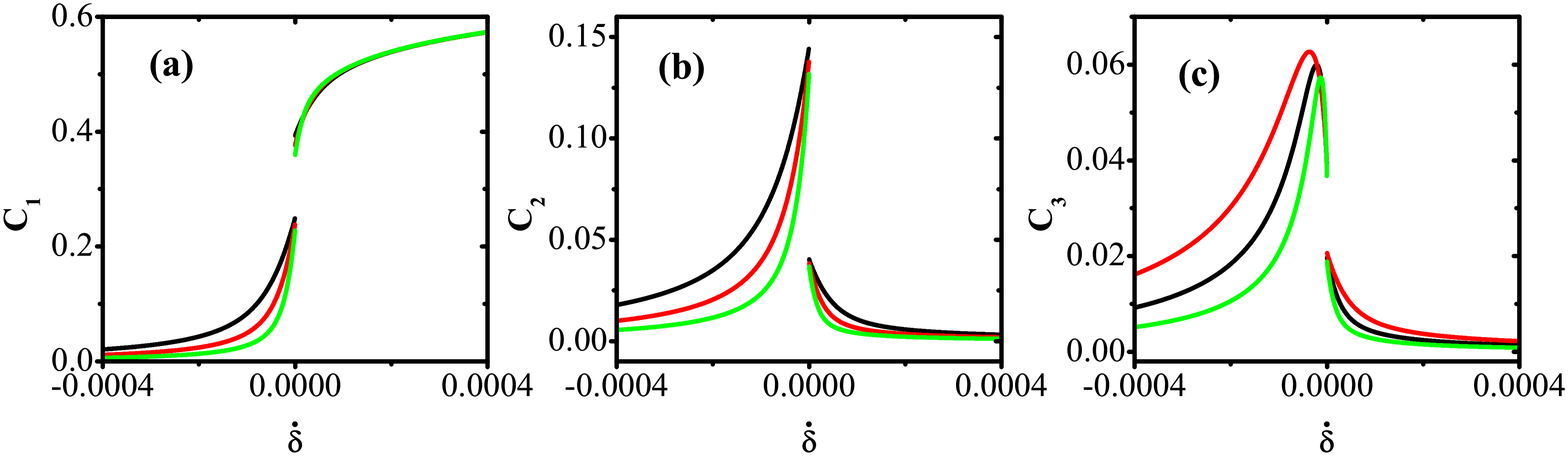}}
\resizebox{0.5\textwidth}{!}{\includegraphics{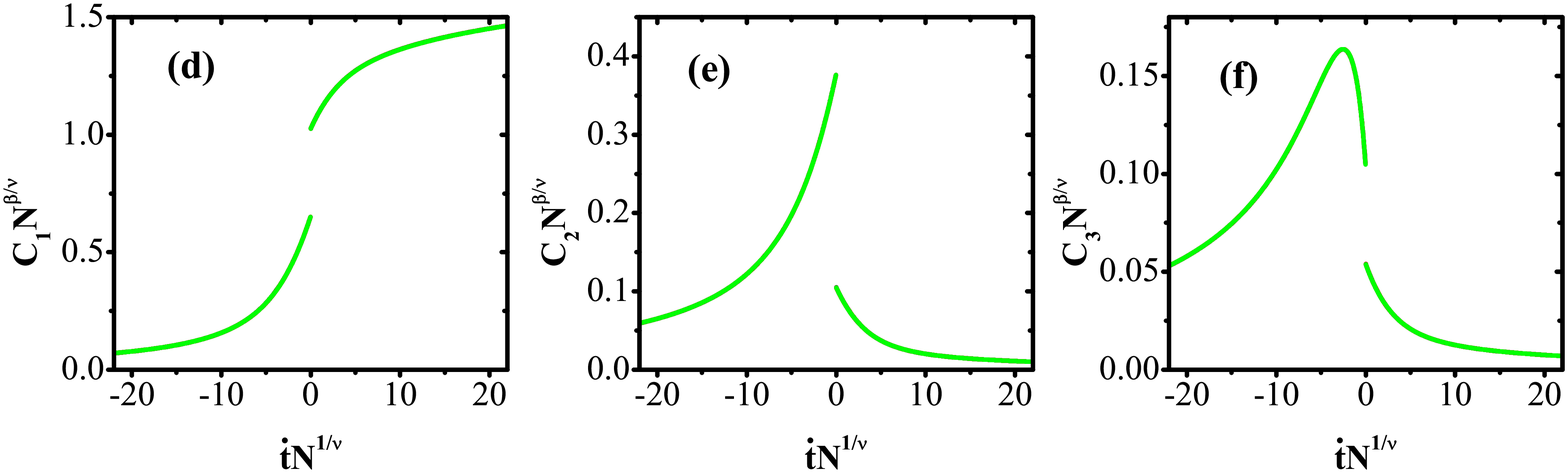}}
\caption{Plots of modified sizes of ranked clusters $C_R$ versus $\dot{\delta}$ and modified FSS functions $\tilde{C}_R=C_{R}N^{\beta/\nu}$ versus scaled variable $\dot{t}N^{1/\nu}$ for three largest clusters of PR model. We take $1/\nu=0.74$ , $\beta/\nu=0.065$. For each $R$ three curves with system sizes $N/10^4=256$(black), 512 (red), 1024 (green) are plotted.}
\label{fig:3}       % Give a unique label
\end{figure}

Without loss of generality, we take PR model as an example. In Fig.\ref{fig:3}(a)(b)(c), the modified ranked clusters $C_1$, $C_2$ and $C_3$ with respect to $\dot{\delta}$ are plotted with $N/10^4=256, 512, 1024$ and the curves of $C_R$ with $R\ge 4$ behave in the similar way as $C_2$ and $C_3$. Due to the way of averaging, the largest gap of largest cluster is revealed as sudden sharp increase of $C_1$ at $\dot{\delta}=0$. Meanwhile, $C_2$ and $C_3$ show sudden sharp decrease. The sudden changes of $C_R$ at $\dot{\delta}=0$ seem to announce the discontinuity of percolation. However, it's only finite size effect because the size jumps in $C_R$ will converge to 0 as $N \to \infty$ as we demonstrated above. In fact, at the pseudo-critical point in each realization of network, the largest cluster merges with the second largest one to become the new largest cluster and the ranks of other clusters all rise by one. Therefore, the gap of $C_1$ at $\dot{\delta}=0$ are actually the $C_2$ right before the jump.

In Fig.\ref{fig:3}(d)(e)(f), we demonstrate the modified FSS functions $\tilde{C}_R=C_{R}N^{\beta/\nu}$ with respect to the scaled variable $\dot{t}N^{1/\nu}$ where $1/\nu=1/\nu_1=0.74$ and $\beta/\nu=\beta_1=0.065$ for PR model. Obviously, data of $C_R$ with different system sizes collapse together almost perfectly and it remains so in the range much larger than we show here. In contrast, as shown in Ref.\cite{Fan12}, in the framework of standard FSS theory, curves of $s_2/s_1$ of PR model with different system sizes separate soon as the scaled variable $tN^{1/\nu}\neq 0$.

\begin{figure}
\resizebox{0.45\textwidth}{!}{\includegraphics{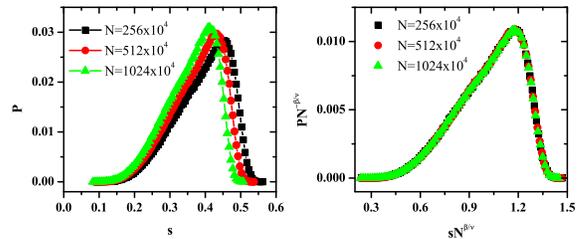}}
\caption{Plots of $P(s,N)$ the distribution of order parameter at the sample-dependent pseudo-critical point where $\dot{\delta}=0$ and the corresponding FSS function $\tilde{P}(sN^{\beta/\nu})=P(s,N)N^{-\beta/\nu}$ with $\beta/\nu=0.065$ for PR model. }
\label{fig:4}       % Give a unique label
\end{figure}

As we mentioned above, the distribution of order parameter at the critical point of infinite system $r_c(\infty)$ exhibits double peaks. That's because $r_c(\infty)$ is larger than the sample-dependent pseudo-critical point $r_c^{i}$ in some realizations while not in the others. In this case, the distribution of order parameter should be measured at the sample-dependent pseudo-critical point where $\dot{\delta}=0$ rather than at $r_c(\infty)$. As presented for PR model in Fig.\ref{fig:4}, distribution $P_{\dot{\delta}=0}(s,N)$ exhibits only one peak and could be perfectly collapsed into a scaling function $\tilde{P}(sN^{\beta/\nu})$ where $\beta/\nu=0.065$.

{\it Summary}---
In summary, we have studied the finite size scaling behaviors of both classic random percolation and explosive percolation in random networks with Monte Carlo simulation.

In the framework of standard FSS theory, sizes of ranked clusters, $s_R$, all converge to zero in power-law with exponent $\beta/\nu$ at the critical point. However, $s_R$ with different $R$ leads to different $\beta/\nu$ in explosive percolation. Moreover, the obtained critical exponents $1/\nu$ and $\beta/\nu$ for CDGM model are not in consistent with the analytical results.

During the process of adding edge one by one to complex networks, the size of largest cluster experiences a series of jumps which are closely related to phase transition. We define the reduced edge number $r$ at which the largest size jump of largest cluster induced by single edge occurs in each realization as sample-dependent pseudo-critical point. From the asymptotic behaviors of sample-dependent pseudo-critical points and size jumps there, we obtain right critical exponents for explosive percolation. Based on this, we propose a modified finite size scaling form where sizes of ranked clusters are averaged at same distances from the sample-dependent pseudo-critical point in each realization instead of averaging at same value of control parameter.

\end{document}